\documentclass[10pt,conference]{IEEEtran}\IEEEoverridecommandlockouts
\usepackage{epsfig,graphicx,subfigure,psfrag,amsmath,cases}
\usepackage{latexsym,amssymb,amsmath,epsfig,subfigure,algorithm,mathtools}
\usepackage{algorithmic}
\usepackage{color}
\usepackage{url}
\usepackage{scrtime}
\usepackage[left=0.6in,top=0.48in,bottom=0.53in,right=0.6in]{geometry}
\author{\authorblockN{Yan Sun\authorrefmark{1}, Derrick Wing Kwan
Ng\authorrefmark{2}, and Robert Schober\authorrefmark{1}\thanks{Derrick
Wing Kwan Ng and Robert Schober are also with the University of British
Columbia, Canada. }}

Institute for Digital Communications, Friedrich-Alexander-University
Erlangen-N\"urnberg (FAU), Germany\authorrefmark{1}\authorrefmark{2}\\
School of Electrical Engineering and Telecommunications, The University
of New South Wales, Australia\authorrefmark{2} \vspace*{-3mm}

}

\title{Multi-Objective Optimization for Power Efficient Full-Duplex
Wireless Communication Systems \vspace*{-1mm}
}

\newtheorem{Thm}{Theorem}

\newtheorem{Prob}{Problem}
\newtheorem{T-Prob}{Transformed Problem}

\DeclareMathOperator{\Tr}{Tr}
\DeclareMathOperator{\Rank}{Rank}
\DeclareMathOperator{\maxo}{maximize}
\DeclareMathOperator{\mino}{minimize}
  \newcommand{\qed}{\hfill \ensuremath{\blacksquare}}
\newtheorem{Remark}{Remark}

\newcommand{\abs}[1]{\lvert#1\rvert}
\newcommand{\norm}[1]{\lVert#1\rVert}
\begin{document}
\maketitle

\begin{abstract}
In this paper, we investigate power efficient resource allocation algorithm design for multiuser wireless communication systems employing a full-duplex (FD) radio base station for serving multiple half-duplex (HD) downlink and uplink users simultaneously. We propose a multi-objective optimization framework for achieving two conflicting yet desirable system design objectives, i.e., total downlink transmit power minimization and total uplink transmit power minimization, while guaranteeing the quality-of-service of all users. To this end, the weighted Tchebycheff method is adopted to formulate a multi-objective optimization problem (MOOP). Although the considered MOOP is non-convex, we solve it optimally by semidefinite programming relaxation. Simulation results not only unveil the trade-off between the total downlink and the total uplink transmit power, but also confirm that the proposed FD system provides substantial power savings over traditional HD systems.
\end{abstract}
\renewcommand{\baselinestretch}{0.979}
\normalsize

\section{Introduction}
\label{sect1}

The exponential growth in high data rate communication has triggered a tremendous demand for radio resources such as bandwidth and energy. A relevant technique for reducing the energy and bandwidth consumption for given quality-of-service (QoS) requirements is multiple-input multiple-output (MIMO) as it offers extra degrees of freedom for more efficient resource allocation. However, the MIMO gain may be difficult to achieve in practice due to the high computational complexity of MIMO receivers \cite{zhang2013gallager}. As an alternative, multiuser MIMO (MU-MIMO) has been proposed as an effective technique for achieving the MIMO performance gain. In particular, in MU-MIMO, a transmitter with multiple antennas (e.g. a base station (BS)) serves multiple single-antenna users which shifts the computational complexity from the receivers to the transmitter \cite{MU_MIMO}. Nevertheless, the spectral resource is still underutilized even if MU-MIMO is employed due to the traditional half-duplex (HD) operation of the BS, where uplink and downlink communication are separated in either time or frequency which leads to a significant loss in spectral efficiency.

As a result, full-duplex (FD) wireless communications has recently received significant attention from both academia and industry due to its potential to double the spectral efficiency of the existing wireless communication systems \cite{FDRad}\nocite{lowcomplex_FDrelay,Dynamic_FDrelay,Mpair_FDRelay_massive}--\cite{FD_smlcll}. In contrast to conventional HD transmission, FD enables simultaneous downlink and uplink transmission at the same frequency. However, in practice, a major challenge in FD communication is self-interference (SI) caused by the signal leakage from the downlink transmission to the uplink signal reception.
Besides, the resource allocation for FD systems is complicated by the co-channel interference (CCI) caused by the uplink transmission to the downlink transmission. Thus, different resource allocation designs for FD systems were proposed  and studied to overcome these challenges. The authors of \cite{lowcomplex_FDrelay} investigated the end-to-end outage probability of MIMO FD single-user relaying systems. In \cite{Dynamic_FDrelay}, a resource allocation algorithm was proposed for the maximization of the end-to-end system data rate of multi-carrier MIMO FD relaying systems. In \cite{Mpair_FDRelay_massive}, massive MIMO was applied in FD relaying systems to facilitate SI suppression and to improve spectral efficiency.  Simultaneous downlink and uplink transmission in small cells was studied in \cite{FD_smlcll}, where a suboptimal downlink beamformer was designed to improve the system throughput. However, the power consumption in FD systems has not been thoroughly studied in the literature so far \cite{lowcomplex_FDrelay}\nocite{Dynamic_FDrelay, Mpair_FDRelay_massive}--\cite{FD_smlcll}.

On the other hand, a large amount of work has been devoted to power efficient/green communications because of the pressure stemming from environmental concerns and the rapidly increasing cost of energy \cite{PE_MISO_Sec}\nocite{MU_ULDL_BF,JPMrelay}--\cite{EE_OFDMA_largeant}.
In \cite{PE_MISO_Sec}, downlink beamforming was proposed to minimize the total transmit power while guaranteeing secure communication for a single desired user.
In \cite{MU_ULDL_BF}, both uplink and downlink beamforming were studied for total transmit power minimization in MU-MIMO systems.
In \cite{JPMrelay}, a power efficient scheduling scheme was designed to minimize the power consumption of multihop wireless links.
In \cite{EE_OFDMA_largeant}, the maximization of the energy efficiency of multi-carrier systems was studied. However, all of the above works focus on HD systems  and the obtained results may not be applicable to FD communication systems.
In particular, total downlink and total uplink transmit power minimization are conflicting design objectives in FD systems. However, the intricate trade-off between downlink and uplink power consumption in FD systems has not been studied in the literature \cite{PE_MISO_Sec}\nocite{MU_ULDL_BF,JPMrelay}--\cite{EE_OFDMA_largeant}. Such a trade-off does not exist in HD systems since downlink and uplink transmission are separated either in the time domain or in the frequency domain.
Besides, the single-objective optimization frameworks in \cite{PE_MISO_Sec}\nocite{MU_ULDL_BF,JPMrelay}--\cite{EE_OFDMA_largeant} are not suitable for studying the trade-off between downlink and uplink power consumption in FD systems.

Motivated by the aforementioned observations, we propose a multi-objective optimization problem (MOOP) formulation which minimizes the total downlink and the total uplink transmit powers subject to QoS constraints for MU-MIMO FD wireless communication systems.
The proposed MOOP is formulated by adopting the weighted Tchebycheff method and solved optimally by semidefinite programming (SDP) relaxation.

\section{System Model}
In this section, we present the adopted MU-MIMO channel model for FD wireless communication systems.

\subsection{Notation}
We use boldface capital and lower case letters to denote matrices and vectors, respectively. $\mathbf{A}^H$, $\Tr(\mathbf{A})$, and $\Rank(\mathbf{A})$ represent the  Hermitian transpose, trace, and rank of  matrix $\mathbf{A}$, respectively; $\mathbf{A}^{-1}$ and $\mathbf{A}^{\dagger}$ represent the inverse and Moore-Penrose pseudoinverse of matrix $\mathbf{A}$, respectively; $\mathbf{A}\succeq \mathbf{0}$ indicates that $\mathbf{A}$ is a positive semidefinite matrix; $\mathbf{I}_N$ is the $N\times N$ identity matrix; $\mathbb{C}^{N\times M}$ denotes the set of all $N\times M$ matrices with complex entries; $\mathbb{H}^N$ denotes the set of all $N\times N$ Hermitian matrices; $\abs{\cdot}$ and $\norm{\cdot}$ denote the absolute value of a complex scalar and the Euclidean vector norm, respectively; ${\cal E}\{\cdot\}$ denotes statistical expectation; the circularly symmetric complex Gaussian distribution with mean $\mu$ and variance $\sigma^2$ is denoted by ${\cal CN}(\mu,\sigma^2)$; and $\sim$ stands for ``distributed as".

\subsection{Multiuser System Model}
\begin{figure}
\centering\vspace*{-0.5cm}
\includegraphics[width=3.5in]{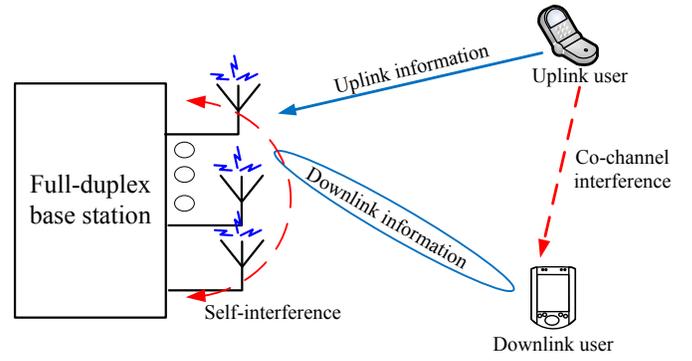}\vspace*{-0.3cm}
\caption{A multiuser communication system with a full-duplex (FD) radio base station (BS), $K=1$ half-duplex (HD) downlink users, and $J=1$ HD uplink users. The BS is equipped with $N_\mathrm{T}$ antennas for simultaneous uplink and downlink communication.}
\label{fig:system_model}\vspace*{-0.3cm}
\end{figure}

We consider a multiuser communication system. The system consists of an FD radio BS, $K$ downlink users, and $J$ uplink users, cf. Figure \ref{fig:system_model}. The FD radio BS is equipped with $N_{\mathrm{T}} > 1$ antennas for simultaneous downlink transmission and uplink reception in the same frequency band\footnote{We note that circulator based FD radio prototypes which can transmit and receive signals simultaneously on the same antennas have been demonstrated \cite{FDRad}.}. The $K+J$ users are single-antenna HD mobile communication devices to ensure low hardware complexity. The downlink and the uplink users are scheduled for simultaneous uplink and downlink transmission. Besides, we assume that the global channel state information (CSI) of all users is perfectly known at the BS for resource allocation.

\subsection{Channel Model}

We focus on a frequency flat fading channel. The received signals at downlink user $k\in\{1,\ldots,K\}$ and the FD radio BS are given by
\begin{eqnarray}
\label{eqn:dl_user_rcv_signal}y_{k}^{\mathrm{DL}}\hspace*{-2mm}&=&\hspace*{-2mm}\mathbf{h}_k^H\mathbf{x}_k\hspace*{3mm}+\hspace*{-0.5mm}\underbrace{\sum_{m\neq k}^K\mathbf{h}_k^H\mathbf{x}_m}_{\mbox{multiuser interference}}\hspace*{3mm} \notag\\
&& + \hspace*{-0.5mm}\underbrace{\sum_{j=1}^J \sqrt{P_j}f_{j,k}d_j^{\mathrm{UL}}}_{\mbox{co-channel interference}}\hspace*{-0.5mm} + \hspace*{3mm} n^{\mathrm{DL}}_{k}\hspace*{-0.5mm},\,\,  \text{and}\\
\label{eqn:ul_rcv_signal}\mathbf{y}^{\mathrm{UL}}\hspace*{-2mm}&=&\hspace*{-2mm}\sum_{j=1}^J \sqrt{P_j}\mathbf{g}_j d_j^{\mathrm{UL}}\hspace*{3mm}+\hspace*{-0.5mm} \underbrace{\mathbf{H}_{\mathrm{SI}}\sum_{k=1}^K{\mathbf{x}_k}}_{\mbox{self-interference}}\hspace*{-0.5mm}+\hspace*{3mm}\mathbf{z},\,\,
\end{eqnarray}
respectively, where $\mathbf{x}_k\in\mathbb{C}^{N_{\mathrm{T}}\times1}$ denotes the transmit signal vector from the FD radio BS to downlink user $k$. The downlink channel between the FD radio BS and user $k$ is denoted by $\mathbf{h}_k\in\mathbb{C}^{N_{\mathrm{T}}\times1}$ and $f_{j,k}\in\mathbb{C}$ represents the channel between uplink user $j$ and downlink user $k$. Variables $d_j^{\mathrm{UL}}$ and $P_j$ are the transmit data and the transmit power sent from uplink user $j$ to the FD radio BS, respectively. Vector $\mathbf{g}_j\in\mathbb{C}^{N_{\mathrm{T}}\times1}$ denotes the channel between uplink user $j$ and the FD radio BS. Matrix $\mathbf{H}_{\mathrm{SI}}\in{\mathbb{C}^{N_{\mathrm{T}}\times N_{\mathrm{T}}}}$ denotes the SI channel at the FD radio BS. Variables $\mathbf{h}_k$, $f_{j,k}$, $\mathbf{g}_j$, and $\mathbf{H}_{\mathrm{SI}}$ capture the joint effect of path loss and small scale fading.  $\mathbf{z}\sim{\cal CN}(0,\sigma_{\mathrm{z}}^2\mathbf{I}_{N_\mathrm{T}})$ and $n^{\mathrm{DL}}_{k}\sim{\cal CN}(0,\sigma_{\mathrm{n}_k}^2)$ represent the additive white Gaussian noise (AWGN) at the FD radio BS and downlink user $k$, respectively.
In (\ref{eqn:dl_user_rcv_signal}), the term $\sum_{m\neq k}^K\mathbf{h}_k^H\mathbf{x}_m$ represents the multiuser interference (MUI) in the downlink channel, and the term $\sum_{j=1}^J \sqrt{P_j}f_{j,k}d_j^{\mathrm{UL}}$ denotes the CCI created by the uplink users to downlink user $k$. In (\ref{eqn:ul_rcv_signal}), the term $\mathbf{H}_{\mathrm{SI}}\sum_{k=1}^K{\mathbf{x}_k}$ represents the SI.

In each scheduling time slot, the FD radio BS transmits $K$ independent signal streams simultaneously at the same frequency to the $K$ downlink users. In particular, the transmit signal vector
\begin{eqnarray}
\mathbf{x}_k=\mathbf{w}_k d_k^{\mathrm{DL}}
\end{eqnarray}
is adopted at the BS, where $d_k^{\mathrm{DL}}\in\mathbb{C}$ and $\mathbf{w}_k\in\mathbb{C}^{N_\mathrm{T}\times1}$ are the information bearing signal for downlink user $k$ and the corresponding beamforming vector, respectively. Without loss of generality, we assume ${\cal E}\{\abs{d_k^{\mathrm{DL}}}^2\}={\cal E}\{\abs{d_j^{\mathrm{UL}}}^2\}=1,\forall k\in\{1,\ldots,K\}, j\in\{1,\ldots,J\}$.

\section{Problem Formulation}\label{sect:forumlation}
In this section, we first define the adopted performance metric for the considered multiuser communication system. Then, we formulate the downlink and uplink power allocation problems. The number of antennas at the FD radio BS is assumed to be larger than the number of uplink users to facilitate uplink signal detection, i.e., $N_\mathrm{T} \ge J$.

\subsection{Performance Metrics}
The receive signal-to-interference-plus-noise ratio (SINR) at downlink user $k$ is given by
\begin{eqnarray}
\Gamma^{\mathrm{DL}}_{k}=\frac{\abs{\mathbf{h}_k^H\mathbf{w}_k}^2}{\overset{K}{\underset{m \neq k}{\sum}}\abs{\mathbf{h}_k^H\mathbf{w}_m}^2 + \overset{J}{\underset{j=1}{\sum}}P_j\abs{f_{j,k}}^2 + \sigma_{\mathrm{n}_k}^2}.
\end{eqnarray}

On the other hand, the receive SINR of uplink user $j$ at the FD radio BS is given by
\begin{eqnarray}
\Gamma^{\mathrm{UL}}_{j}=\frac{P_j\abs{\mathbf{g}_j^H\mathbf{v}_j}^2}{\overset{J}{\underset{r \neq j}{\sum}}P_r\abs{\mathbf{g}_r^H\mathbf{v}_j}^2+\overset{K}{\underset{k=1}{\sum}}\abs{\mathbf{v}_j^H\mathbf{H}_{\mathrm{SI}}\mathbf{w}_k}^2 +\sigma_{\mathrm{z}}^2\norm{\mathbf{v}_j}^2},
\end{eqnarray}
where $\mathbf{v}_j\in\mathbb{C}^{N_{\mathrm{T}}\times1}$ is the receive beamforming vector for decoding of the information received from uplink user $j$. In this paper, zero-forcing receive beamforming (ZF-BF) is adopted. We note that ZF-BF closely approaches the performance of optimal minimum mean square error beamforming (MMSE-BF) when the noise term is not dominating \cite{book:david_wirelss_com}\footnote{We note that the noise power at the BS is not expected to be the dominating factor for the system performance since the BS is usually equipped with a high quality low-noise amplifier (LNA).} or the number of antennas is sufficiently large \cite{Mpair_FDRelay_massive}. Besides, ZF-BF facilitates the design of a computational efficient resource allocation algorithm.
Hence, the receive beamformer for uplink user $j$ is chosen as
 \begin{eqnarray}
 \mathbf{v}_j=(\mathbf{u}_j\mathbf{Q}^{\dagger})^H ,
  \end{eqnarray}
 where $\mathbf{u}_j=\Big[\underbrace{0,\ldots,0}_{(j-1)},1,\underbrace{0,\ldots,0}_{(J-j)}\Big]$, $\mathbf{Q}^{\dagger}=(\mathbf{Q}^H \mathbf{Q})^{-1}\mathbf{Q}^H$, and $\mathbf{Q}=[\mathbf{g}_1,\ldots,\mathbf{g}_J]$.

\subsection{Optimization Problem Formulation}
\label{sect:cross-Layer_formulation}
We first study the problem formulation of two desirable system design objectives for the considered FD wireless system. Then, we investigate the two system objectives jointly under the framework of multi-objective optimization. The first considered problem is designed to minimize the total downlink transmit power and given by
\begin{Prob}[Total Downlink Transmit Power Minimization]\end{Prob}\vspace*{-0.88cm}
\begin{eqnarray}
\label{eqn:prob1}
&&\hspace*{-1mm} \underset{\mathbf{w}_k,P_j}{\mino}\,\, \,\, \sum_{k=1}^{K}\norm{\mathbf{w}_k}^2 \notag \vspace*{-0.5cm}\\\vspace*{-0.5cm}
\notag\mbox{s.t.}
&&\hspace*{-5mm}\mbox{C1: }\frac{\abs{\mathbf{h}_k^H\mathbf{w}_k}^2}{\overset{K}{\underset{m \neq k}{\sum}}\abs{\mathbf{h}_k^H\mathbf{w}_m}^2 + \overset{J}{\underset{j=1}{\sum}}P_j\abs{f_{j,k}}^2 + \sigma_{\mathrm{n}_k}^2} \geq \Gamma^{\mathrm{DL}}_{\mathrm{req}_k},\,\, \forall k, \notag\\
&&\hspace*{-5mm}\mbox{C2: }\frac{P_j\abs{\mathbf{g}_j^H\mathbf{v}_j}^2}{S^{\mathrm{UL}}_j +\sigma_{\mathrm{z}}^2\norm{\mathbf{v}_j}^2} \geq \Gamma^{\mathrm{UL}}_{\mathrm{req}_j},\,\, \forall j, \notag\\
&&\hspace*{-5mm}\mbox{C3: } P_j \geq 0,\,\, \forall j,
\end{eqnarray}
where we define $S^{\mathrm{UL}}_j=\overset{J}{\underset{r \neq j}{\sum}}P_r\abs{\mathbf{g}_r^H\mathbf{v}_j}^2+\overset{K}{\underset{k=1}{\sum}}
\abs{\mathbf{v}_j^H\mathbf{H}_{\mathrm{SI}}\mathbf{w}_k}^2$ for notational simplicity. Besides, $\Gamma^{\mathrm{DL}}_{\mathrm{req}_k} > 0$ and $\Gamma^{\mathrm{UL}}_{\mathrm{req}_j} > 0$ are the minimum required SINRs for downlink user $k\in\{1,\ldots,K\}$ and uplink user $j\in\{1,\ldots,J\}$, respectively.

Constraint C3 in (\ref{eqn:prob1}) is the non-negative power constraint for uplink user $j$. Problem $1$ is a non-convex problem due to constraints C1 and C2. We note that Problem $1$ tries to minimize only the total downlink transmit power under constraints C1--C3 without regard for the consumed uplink transmit powers.

The second desirable system objective is the minimization of total uplink transmit power and can be mathematically formulated as
\begin{Prob}[Total Uplink Transmit Power Minimization]\end{Prob}\vspace*{-0.4cm}
\begin{eqnarray}
\label{eqn:prob2}
&&\hspace*{-15mm} \underset{\mathbf{w}_k,P_j}{\mino}\,\, \,\, \sum_{j=1}^{J}P_j \notag\\
\mbox{s.t.} &&\hspace*{-3mm}\mbox{C1 -- C3}.
\end{eqnarray}
Problem $2$ targets only the minimization of the total uplink transmit power under constraints C1--C3 without regard for the consumed downlink transmit power.

The two above system design objectives are desirable for the system operator and the users, respectively. However, in FD wireless communication systems, these  objectives conflict with each other.
On the one hand, the uplink transmission causes CCI to the downlink transmission. Hence, the FD radio BS should transmit with a high power to ensure the QoS requirements of the downlink users. On the other hand, a high downlink transmit power results in strong SI to the uplink reception. Hence, the uplink users should transmit with a high power to satisfy the minimum required receive SINRs of the uplink users at the FD radio BS. However, this would in turn cause high CCI and give rise to successive increases in transmit power for both uplink and downlink transmission. To overcome this problem, we resort to multi-objective optimization. In the literature, multi-objective optimization is often adopted to study the trade-off between conflicting system design objectives via the concept of Pareto optimality. A point is Pareto optimal if there is no other point that improves at least one of the objectives without detriment to another objective\footnote{Please refer to \cite{JR:MOOP} for a detailed discussion of Pareto optimality.}. In order to capture the complete Pareto optimal set, we formulate a third optimization problem to investigate the trade-off between Problem $1$ and Problem $2$ by using the weighted Tchebycheff method \cite{JR:MOOP}. The resulting problem formulation is given as:
\begin{Prob}[Multi-Objective Optimization]\end{Prob}\vspace*{-0.37cm}
\begin{eqnarray}
\label{eqn:MOP}&&\hspace*{-10mm}\underset{\mathbf{w}_k,P_j}{\mino}\,\,\max_{i=1,2}\,\, \Big\{\lambda_i \Big(Q_i(\mathbf{w}_k,P_j)-Q_i^*\Big)\Big\}\nonumber\\
&&\hspace*{6mm}\mbox{s.t.}\hspace*{4mm} \mbox{C1 -- C3},
\end{eqnarray}
where $Q_1(\mathbf{w}_k,P_j)=\sum_{k=1}^{K}\norm{\mathbf{w}_k}^2$ and $Q_2(\mathbf{w}_k,P_j)=\sum_{j=1}^{J}P_j$.
Constant $Q_i^*$ is the optimal objective value of the $i$-th problem. In particular, variable $Q_i^*$ is assumed to be a constant for resource allocation. Variable $\lambda_i\ge 0$, $\sum_i\lambda_i=1$, specifies the priority of the $i$-th objective compared to the other objectives and reflects the preference of the system operator. By varying $\lambda_i$, we can obtain a complete Pareto optimal set which corresponds to a set of resource allocation policies. Thus, the operator can select a proper resource allocation policy from the set of available policies. Compared to other formulation methods for handing MOOPs in the literature (e.g. the weighted product method and the exponentially weighted criterion \cite{JR:MOOP}), the weighted Tchebycheff method can achieve the complete Pareto optimal set with a lower computational complexity, despite the non-convexity (if any) of the considered problem. It is noticed that Problem $3$ is equivalent to Problem $i$ when $\lambda_i=1$ and $\lambda_j=0$,  $\forall i\ne j$. Here, equivalent means that both problem formulations have the same optimal solution.
\section{Solution of the Optimization Problems} \label{sect:solution}
The optimization problems in (\ref{eqn:prob1}), (\ref{eqn:prob2}), and (\ref{eqn:MOP}) are non-convex problems due to the non-convexity of constraints C1 and C2. In order to solve the problems efficiently, we recast Problems $1$, $2$, and $3$ as convex optimization problems via SDP relaxation.
\subsection{Semidefinite Programming Relaxation} \label{sect:solution_dual_decomposition}
For facilitating the SDP relaxation, we define
\begin{eqnarray}\label{eqn:change_of_variables}
\mathbf{W}_k=\mathbf{w}_k\mathbf{w}_k^H,\,\mathbf{H}_k=\mathbf{h}_k\mathbf{h}_k^H,\,\mathbf{G}_j=\mathbf{g}_j\mathbf{g}_j^H,\,\mathbf{V}_j=\mathbf{v}_j\mathbf{v}_j^H,
\end{eqnarray}
and rewrite Problems $1$-$3$ into the following equivalent formulations:

\begin{T-Prob}\end{T-Prob} \vspace*{-0.5cm}
\begin{eqnarray}
\label{eqn:sdp1}&&\hspace*{1mm}\underset{{\mathbf{W}_k}\in \mathbb{H}^{N_{\mathrm{T}}},P_j
}{\mino}\,\, \sum_{k=1}^{K}\Tr(\mathbf{W}_k)\nonumber\\
\notag \mbox{s.t.} &&\hspace*{-5mm}\widetilde{\text{C1}}\mbox{: } \notag\frac{\Tr(\mathbf{H}_k\mathbf{W}_k)}{\Gamma^{\mathrm{DL}}_{\mathrm{req}_k}} \ge {I_k^{\mathrm{DL}}+\sigma_{{\mathrm{n}}_k}^2},\,\, \forall k, \\
&&\hspace*{-5mm}\widetilde{\text{C2}}\mbox{: }\notag\frac{P_j\Tr(\mathbf{V}_j\mathbf{G}_j)}{\Gamma^{\mathrm{UL}}_{\mathrm{req}_j}}
 \ge {I_j^\mathrm{UL}+\sigma_{\mathrm{z}}^2\Tr(\mathbf{V}_j)},\,\, \forall j,\\
&&\hspace*{-5mm}\widetilde{\text{C3}}\mbox{: }\,\, P_j \geq 0,\,\, \forall j,\notag\\
&&\hspace*{-5mm}\widetilde{\text{C4}}\mbox{: }\,\, \mathbf{W}_k\succeq \mathbf{0},\,\, \forall k,\notag\\
&&\hspace*{-5mm}\widetilde{\text{C5}}\mbox{: }\,\, \Rank(\mathbf{W}_k) \le 1,\,\, \forall k,
\end{eqnarray}
where $\mathbf{W}_k\succeq \mathbf{0}$, ${\mathbf{W}_k}\in \mathbb{H}^{N_\mathrm{T}}$, and $\Rank(\mathbf{W}_k) \le 1$ in (\ref{eqn:sdp1}) are imposed to guarantee that $\mathbf{W}_k=\mathbf{w}_k\mathbf{w}_k^H$ holds after optimization, and
\begin{eqnarray}
I_k^\mathrm{DL}\hspace*{-2mm}&=&\hspace*{-2mm} \overset{K}{\underset{m \neq k}{\sum}} \Tr(\mathbf{H}_k
\mathbf{W}_m)+\overset{J}{\underset{j=1}{\sum}} P_j\abs{f_{j,k}}^2,\,\, \text{and}\\
I_j^\mathrm{UL}\hspace*{-2mm}&=&\hspace*{-2mm}\overset{J}{\underset{r \neq j}{\sum}} P_r\Tr(\mathbf{V}_j
\mathbf{G}_r)+\overset{K}{\underset{k=1}{\sum}} \Tr(\mathbf{W}_k\mathbf{H}_{\mathrm{SI}}^H\mathbf{V}_j\mathbf{H}_{\mathrm{SI}}).
\end{eqnarray}

\begin{T-Prob}\end{T-Prob} \vspace*{-0.36cm}
\begin{eqnarray}
\label{eqn:sdp2}&&\hspace*{-10mm}\underset{{\mathbf{W}_k}\in \mathbb{H}^{N_{\mathrm{T}}},P_j
}{\mino}\,\, \sum_{j=1}^{J}P_j\nonumber\\
&&\hspace*{-14mm}\mbox{s.t.  }\hspace*{4mm}\widetilde{\text{C1}}\mbox{ -- }\widetilde{\text{C5}}.
\end{eqnarray}

\begin{T-Prob}\end{T-Prob} \vspace*{-0.36cm}
\begin{eqnarray}
\label{eqn:sdp_MOP}&&\hspace*{3mm}\underset{{\mathbf{W}_k}\in \mathbb{H}^{N_{\mathrm{T}}},P_j,t
}{\mino}\,\,\,t \nonumber\\
&&\hspace*{-9mm}\notag \mbox{s.t.}\hspace*{12mm}\widetilde{\text{C1}}\mbox{ -- }\widetilde{\text{C5}}, \notag \\
&&\hspace*{-10mm}\widetilde{\text{C6}}\mbox{: }\lambda_i (Q_i-Q_i^*)\le t, \forall i\in\{1,2\},
\end{eqnarray}
where $t$ is an auxiliary optimization variable and (\ref{eqn:sdp_MOP}) is the epigraph representation of (\ref{eqn:MOP}).

Since Problem $3$ is a generalization of Problems $1$ and $2$, we focus on the solution of Problem $3$. Transformed Problem $3$ is a non-convex problem due to the rank-one constraint $\widetilde{\text{C5}}$ and solving such a rank-constrained problem is known to be NP-hard \cite{gartner2012approximation}. Hence, we relax constraint $\widetilde{\text{C5}}\mbox{: }\Rank(\mathbf{W}_k) \le 1$ by removing it from the problem formulation, such that the considered problem becomes a convex SDP given by
\begin{eqnarray}
\label{eqn:sdp_relaxation}&&\hspace*{-10mm}\underset{{\mathbf{W}_k}\in \mathbb{H}^{N_{\mathrm{T}}},P_j,t
}{\mino}\,\, t \nonumber\\
&&\hspace*{-15mm}\mbox{s.t.  }\hspace*{4mm}\widetilde{\text{C1}}\mbox{ -- }\widetilde{\text{C4}},\widetilde{\text{C6}}.
\end{eqnarray}
The relaxed convex problem in (\ref{eqn:sdp_relaxation}) can be solved efficiently by standard convex program solvers such as CVX \cite{website:CVX}. Besides, if the solution obtained for a relaxed SDP problem is a rank-one matrix, i.e., $\Rank(\mathbf{W}_k) = 1$ for $\mathbf{W}_k\ne\mathbf{0}, \, \forall k$, then it is also the optimal solution of the original problem. Otherwise, in general, the solution of (\ref{eqn:sdp_relaxation}) serves as a lower bound for the original problem (\ref{eqn:sdp_MOP}) due to the larger feasible solution set in (\ref{eqn:sdp_relaxation}).

Next, we reveal a sufficient condition for obtaining a rank-one solution $\mathbf{W}_k$ for the relaxed version of Transformed Problem $3$ in the following theorem.

\begin{Thm}\label{thm:rankone_condition} Under the condition that the channel vectors of the downlink users, $\mathbf{h}_k,k\in\{1,\ldots,K\},$ and the uplink users, $\mathbf{g}_j,j\in\{1,\ldots,J\},$ can be modeled as statistically independent random variables, the solution of (\ref{eqn:sdp_relaxation}) is rank-one, i.e.,  $\Rank(\mathbf{W}_k)=1$ for $\mathbf{W}_k\ne\mathbf{0}, \, \forall k$, with probability one.
\end{Thm}

\emph{\quad Proof: }Please refer to the Appendix. \qed

In other words, whenever the channels satisfy the condition stated in Theorem $1$, the optimal beamformer $\mathbf{w}^*_k$ of (\ref{eqn:MOP}) can be obtained with probability one.

\begin{Remark}
Exploiting Theorem $1$, Problem $1$ and Problem $2$ can also be solved optimally via SDP relaxation. In particular, for $\lambda_1=1,\lambda_2=0$, and $Q_1^*=0$, the problem in (\ref{eqn:sdp_relaxation}) becomes Problem $1$; for $\lambda_2=1,\lambda_1=0$, and $Q_2^*=0$, the problem in (\ref{eqn:sdp_relaxation}) becomes Problem $2$. We note that setting $Q_i^*=0$
 in Problem $3$ is used only for recovering the solution of Problems $1$ and $2$. However, this does not imply that the optimal value of Problem $i,\forall i\in\{1,2\},$ is equal to $0$.
\end{Remark}

\section{Results}
In this section, we investigate the performance of the proposed multi-objective optimization based resource allocation scheme through simulations.  The simulation parameters are specified in Table \ref{tab:parameters}. There are $K=3$ downlink users and $J=8$ uplink users in the cell. The users are  randomly and uniformly distributed between the reference distance and the  maximum service distance of $250$ meters. The FD radio BS is located at the center of the cell and equipped with $N_\mathrm{T}$ antennas. The small scale fading of the downlink channels, uplink channels, and inter-user channels is modeled as independent and identically distributed Rayleigh fading. The multipath fading coefficients of the SI channel are generated as independent and identically distributed Rician random variables with Rician factor $5$ dB.

\begin{table}[t]\caption{System parameters.}\label{tab:parameters} 
\newcommand{\tabincell}[2]{\begin{tabular}{@{}#1@{}}#2\end{tabular}}
\centering
\begin{tabular}{|l|l|}\hline
\hspace*{-1mm}Carrier center frequency & \mbox{$1.9$ GHz}  \\
\hline
\hspace*{-1mm}System bandwidth & \mbox{$200$ KHz}  \\
\hline
\hspace*{-1mm}Path loss exponent &  \mbox{$3.6$}  \\
\hline
\hspace*{-1mm}Reference distance &  \mbox{$30$ m}  \\
\hline
\hspace*{-1mm}SI cancellation    &  \mbox{$-80$ dB} \cite{FDRad}   \\
\hline
\hspace*{-1mm}Downlink user noise power, $\sigma_{\mathrm{n}_k}^2$ &  \mbox{$-83$ dBm}   \\
\hline
\hspace*{-1mm}Uplink BS noise power, $\sigma_{\mathrm{z}}^2$ &  \mbox{$-110$ dBm}\footnotemark{}   \\
\hline
\hspace*{-1mm}BS antenna gain &  \mbox{$10$ dBi}  \\
\hline
\end{tabular} 
\end{table}

\footnotetext{The downlink user noise power is higher than the uplink BS noise power since, unlike cheap mobile devices, the BS is usually equipped with a high quality LNA and a high resolution analog-to-digital converter (ADC) \cite{holma2009lte}.}

\subsection{Transmit Power Trade-off Region}
\begin{figure}[t]
 \centering
\includegraphics[width=3.5in]{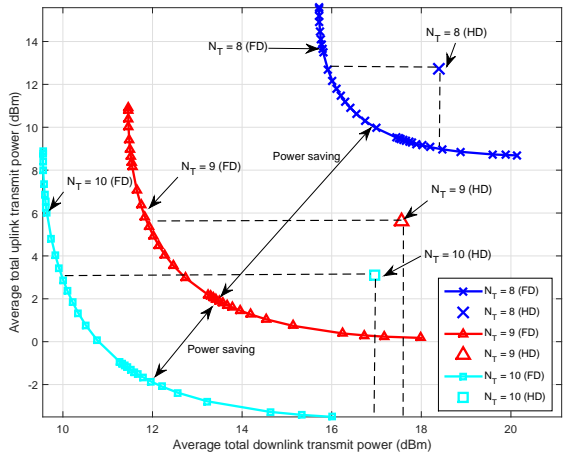}
\caption{Average system objective trade-off region achieved by the proposed optimal resource allocation scheme. The double-sided arrows indicate the power saving due to additional antennas.} \label{fig:trade-off}
\end{figure}

In Figure \ref{fig:trade-off}, we study the trade-off between the downlink and the uplink total transmit powers for different numbers of antennas at the FD radio BS. The trade-off region is obtained by solving (\ref{eqn:sdp_relaxation}) for different $0\leq \lambda_i \leq 1, \forall i \in \{1,2\}$, i.e., the $\lambda_i$ are varied uniformly using a step size of $0.01$ subject to $\sum_i \lambda_i=1$. Without loss of generality, we assume all downlink users and all uplink users require the same SINRs, respectively, i.e., $\Gamma^{\mathrm{DL}}_{\mathrm{req}_k}=\Gamma^{\mathrm{DL}}_{\mathrm{req}}=10$ dB and $\Gamma^{\mathrm{UL}}_{\mathrm{req}_j}=\Gamma^{\mathrm{UL}}_{\mathrm{req}}=6$ dB. It can be observed from Figure 2 that the total uplink transmit power is a monotonically decreasing function with respect to the total downlink transmit power. In other words, minimizing the total uplink power consumption leads to a higher power consumption in the downlink and vice versa. This result confirms that the minimization of the total uplink transmit power and the total downlink transmit power are conflicting objectives. For the case of $N_\mathrm{T}=10$, $10.9$ dB in uplink transmit power can be saved by increasing the total downlink transmit power by $6.5$ dB.
In addition, Figure \ref{fig:trade-off} also indicates that a significant amount of power can be saved in the FD system by increasing the number of antennas. This is due to the fact that the extra degrees freedom offered by the additional antennas facilitate a more power efficient resource allocation. However, there is a diminishing return in the power saving as the numbers of antennas
 at the FD BS increase due to channel hardening \cite{book:david_wirelss_com}.

For comparison, we also consider a baseline multiuser system with a HD radio BS. In particular, the HD radio BS is equipped with $N_\mathrm{T}$ antennas which are utilized for transmitting or receiving in non-overlapping equal-length time intervals. As a result, both SI and CCI are avoided. In order to have a fair comparison between the FD and the HD systems, we set $\log_2(1+\Gamma^{\mathrm{DL}}_{\mathrm{req}_k})=1/2\log_2(1+\Gamma^{\mathrm{DL-HD}}_{\mathrm{req}_k})$ and $\log_2(1+\Gamma^{\mathrm{UL}}_{\mathrm{req}_j})=1/2\log_2(1+\Gamma^{\mathrm{UL-HD}}_{\mathrm{req}_j})$. Thus, for the HD radio BS, the required SINRs for the downlink and uplink users are $\Gamma^{\mathrm{DL-HD}}_{\mathrm{req}_k}=(1+\Gamma^{\mathrm{DL}}_{\mathrm{req}_k})^2-1$, and $\Gamma^{\mathrm{UL-HD}}_{\mathrm{req}_j}=(1+\Gamma^{\mathrm{UL}}_{\mathrm{req}_j})^2-1$, respectively. In addition, the power consumption of downlink and uplink transmission for the HD system is divided by two since either uplink or downlink transmission is performed at a given time. Then, we optimize $\mathbf{W}_k$ to minimize the downlink total transmit power, and we optimize $P_j$ to minimize the uplink total transmit power for the optimal MMSE receiver at the HD radio BS \cite{book:david_wirelss_com}. The power consumption for the baseline HD system is represented by a single point in Figure \ref{fig:trade-off} for a given number of antennas. It can be observed that the power consumption of the baseline system lies above the curve of the corresponding FD system. In the regions which are spanned by the dashed lines and the corresponding FD system performance curves, the FD systems are more power efficient than the corresponding HD systems in both uplink and downlink. In particular, for the case of $N_\mathrm{T}=10$, the FD system facilitates more than $6$ dB power saving for downlink transmission and more than $5$ dB power saving for uplink transmission compared to the corresponding HD system.
\subsection{Average Transmit Power}
\begin{figure}[t]
 \centering
\includegraphics[width=3.5in]{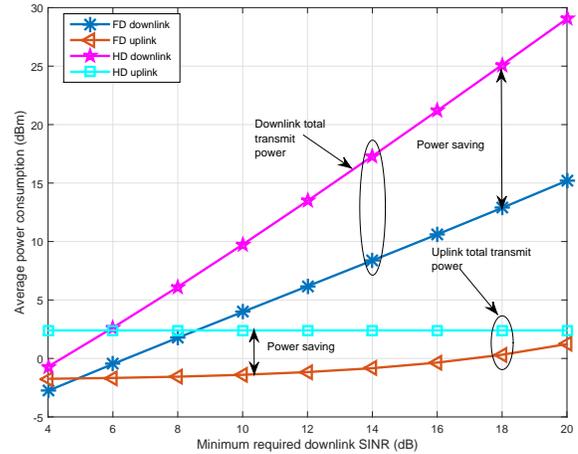} 
\caption{Average power consumption (dBm) versus the minimum required downlink SINRs (dB) for FD and HD radio BS.} \label{fig:figure_power_sinr}
\end{figure}

In Figure \ref{fig:figure_power_sinr}, we investigate the power consumption of downlink and uplink users for identical minimum required downlink SINRs, $\Gamma^{\mathrm{DL}}_{\mathrm{req}_k}=\Gamma^{\mathrm{DL}}_{\mathrm{req}}$, and identical minimum required uplink SINRs of $\Gamma^{\mathrm{UL}}_{\mathrm{req}_j}=6$ dB. The FD radio BS and the baseline HD radio BS are both equipped with $N_\mathrm{T}=10$ antennas. We select the resource allocation policy with $\lambda_1=0.1$ and $\lambda_2=0.9$ which indicates that the system operator attaches a higher priority to total uplink transmit power minimization. It can be observed that both the downlink and the uplink power consumption increase with $\Gamma^{\mathrm{DL}}_{\mathrm{req}}$. However, the downlink power consumption grows more rapidly than the uplink power consumption. The reason behind this is twofold. First, a higher downlink transmit power is required to fulfill more stringent downlink QoS requirements. Second, the SI becomes more severe for higher downlink transmit powers.
Specifically, since the FD radio BS tries to keep the total uplink transmit power low to avoid strong CCI for increasing downlink transmit power, more degrees of freedom at the FD radio BS have to be utilized to suppress the SI to improve the receive SINR of the uplink users. As a result, fewer degrees of freedom are available for reducing the downlink transmit power consumption causing even higher downlink transmit powers. Besides, Figure 3 also confirms that both downlink and uplink transmission in the proposed FD system are more power efficient compared to the HD system.

\section{Conclusions}\label{sect:conclusion}
In this paper, we studied the power allocation algorithm design for MU-MIMO wireless communication systems with an FD BS. The algorithm design was formulated as a non-convex MOOP employing the weighted Tchebycheff method. The proposed MOOP was solved optimally by SDP relaxation. Simulation results not only unveiled the trade-off between the total downlink and the total uplink transmit powers, but also confirmed that the proposed FD system enables substantial power savings compared to a HD system.

\section*{Appendix - Proof of Theorem 1}
The relaxed version of Transformed Problem $3$ in (\ref{eqn:sdp_relaxation}) is jointly convex with respect to the optimization variables and satisfies the Slater's constraint qualification. Therefore, strong duality holds and solving the dual problem is equivalent to solving the primal problem \cite{book:convex}. For obtaining the dual problem, we first need the Lagrangian function of the primal problem in (\ref{eqn:sdp_relaxation}) which is given by
\begin{eqnarray}
&&\hspace*{-6mm}{\cal L}\Big(\mathbf{W}_k,P_j,t,\xi_k,\psi_j,\mathbf{Y}_k,\alpha_j,\theta_i\Big)\\[-0.0cm]
\hspace*{-1mm}=&&\hspace*{-5mm}\sum_{k=1}^K\Tr(\mathbf{R}_k\mathbf{W}_k)-\sum_{k=1}^K\Tr\Big(\mathbf{W}_k
\big(\mathbf{Y}_k+\frac{\xi_k\mathbf{H}_k}{\Gamma^{\mathrm{DL}}_{\mathrm{req}_k}}\big)\Big)+\Delta,\notag
\end{eqnarray}
where
\begin{eqnarray}
\hspace*{-2mm}\Delta\hspace*{-1.5mm}&=&\hspace*{-1.5mm}\sum_{k=1}^K \xi_k\Big(\sigma_{\mathrm{n}_k}^2+\sum_{j=1}^J P_j\abs{f_{j,k}}^2\Big)-\sum_{j=1}^J \alpha_j P_j+\Psi\\
\hspace*{-2mm}&+& \hspace*{-1.5mm}\sum_{j=1}^J \psi_j\Big(\sigma_{\mathrm{z}}^2\Tr(\mathbf{V}_j)+\overset{J}{\underset{r \neq j}{\sum}} P_r\Tr(\mathbf{V}_j\mathbf{G}_r)-\frac{P_j\Tr(\mathbf{V}_j\mathbf{G}_j)}{\Gamma^{\mathrm{UL}}_{\mathrm{req}_j}}\Big)\hspace*{-0.5mm},\notag
\end{eqnarray}
\begin{eqnarray}
\hspace*{-1mm}\Psi\hspace*{-1.5mm}&=&\hspace*{-1.5mm}t(1-\sum_{i=1}^2 \theta_i)-\theta_1\lambda_1 Q_1^*+\theta_2\lambda_2(\sum_{j=1}^J P_j-Q_2^*),\\
\label{eqn:A_k} \hspace*{-2mm}\mathbf{R}_k\hspace*{-1.5mm}&=&\hspace*{-1.5mm}\sum_{m\neq k}^K\xi_m\mathbf{H}_m+\sum_{j=1}^J\psi_j\mathbf{H}_\mathrm{SI}^H\mathbf{V}_j\mathbf{H}_\mathrm{SI}+\theta_1\lambda_1\mathbf{I}_{N_\mathrm{T}}.
\end{eqnarray}
Here, $\Delta$ denotes the collection of terms that only involve variables that are independent of $\mathbf{W}_k$. $\xi_k$, $\psi_j$, $\alpha_j$, and $\theta_i$ are the Lagrange multipliers associated with constraints $\widetilde{\text{C1}}$, $\widetilde{\text{C2}}$, $\widetilde{\text{C3}}$, and $\widetilde{\text{C6}}$. Matrix $\mathbf{Y}_k \in {\mathbb{C}^{N_{\mathrm{T}}\times N_{\mathrm{T}}}} $ is the Lagrange multiplier matrix for the positive semidefinite constraint $\widetilde{\text{C4}}$ on matrix $\mathbf{W}_k$. Thus, the dual problem for the SDP relaxed problem in (\ref{eqn:sdp_relaxation}) is given by
\begin{eqnarray}\label{eqn:dual}
\underset{\underset{\theta_i,\mathbf{Y}_k\succeq \mathbf{0}}{\xi_k,\psi_l,\alpha \ge 0}}{\maxo} \,\,\underset{\underset{P_j,t}{\mathbf{W}_k\in\mathbb{H}^{N_{\mathrm{T}}}}}{\mino} \,\,{\cal L} \Big(\hspace*{-0.5mm}\mathbf{W}_k,P_j,t,\xi_k,\psi_j,\mathbf{Y}_k,\alpha_j,\theta_i\hspace*{-0.5mm}\Big)\notag\\
&&\hspace*{-60mm}\mbox{s.t.}\hspace*{3mm} \sum_{i=1}^2 \theta_i = 1.
\end{eqnarray}
The constraint $\sum_{i=1}^2 \theta_i = 1$ is imposed to guarantee a bounded solution of the dual problem \cite{book:convex}.

We reveal the structure of the optimal $\mathbf{W}_k$ of (\ref{eqn:sdp_relaxation}) by studying the Karush-Kuhn-Tucker (KKT) conditions. For convenience, we define $\mathbf{\Omega}^*\triangleq\{\mathbf{W}^*_k,P^*_j,t^*\}$ and $\mathbf{\Phi}^*\triangleq\{\xi^*_k,\psi^*_j,\mathbf{Y}^*_k,\alpha^*_j,\theta^*_i\}$ as the sets of the optimal primal and dual variables of (\ref{eqn:dual}). In the following, we focus on the KKT conditions which are needed for the proof:
\begin{eqnarray}
\hspace*{-3mm}\mathbf{Y}_k^*\hspace*{-3mm}&\succeq&\hspace*{-3mm}\mathbf{0},\,\,\xi_k^*,\,\psi_j^*,\theta_i^*\ge 0,\,\forall k,\,\forall j,\,\forall i, \label{eqn:dual_variables}\\
\hspace*{-3mm}\mathbf{Y}_k^*\mathbf{W}_k^*\hspace*{-3mm}&=&\hspace*{-3mm}\mathbf{0},\label{eqn:KKT-complementarity}\,\, \text{and}\\
\hspace*{-3mm}\mathbf{Y}_k^*\hspace*{-3mm}&=&\hspace*{-3mm}\mathbf{R}_{k}^*-\xi_k^*\frac{\mathbf{H}_k}{\Gamma^{\mathrm{DL}}_{\mathrm{req}_k}},\label{eqn:lagrangian_gradient}
\end{eqnarray}
where $\mathbf{R}_{k}^*$ is obtained by substituting the optimal dual variables $\mathbf{\Phi}^*$ into (\ref{eqn:A_k}). By exploiting (\ref{eqn:KKT-complementarity}), we have $\mathbf{Y}_k^*\mathbf{W}_k^*=\mathbf{0} \Rightarrow \Rank(\mathbf{Y}_k^*)+\Rank(\mathbf{W}_k^*) \le N_{\mathrm{T}}$. By exploiting the fact that $\Rank(\mathbf{W}_k^*) \ge 1$ for $ \mathbf{W}^*_k\ne\mathbf{0}$, we have
\begin{eqnarray}\label{eqn:rank less than}
\Rank(\mathbf{Y}^*_k) \leq N_{\mathrm{T}}-1.
\end{eqnarray}
If $\Rank(\mathbf{Y}^*_k)=N_{\mathrm{T}}-1$, then the optimal beamforming matrix $\mathbf{W}^*_k\ne \mathbf{0}$ must be a rank-one matrix. In order to further reveal the structure of $\mathbf{Y}^*_k$, we first show by contradiction that $\mathbf{R}_k^*$ is a positive-definite matrix with probability one under the condition stated in Theorem $1$.
Let us focus on the dual problem in (\ref{eqn:dual}). For a given set of optimal dual variables, $\mathbf{\Phi}^*$, the optimal uplink transmit power, $P_j^*$, and the optimal auxiliary variable $t^*$, the dual problem in (\ref{eqn:dual}) can be written as
\begin{eqnarray}\label{eqn:dual2}
\underset{\mathbf{W}_k\in\mathbb{H}^{N_{\mathrm{T}}}}{\mino} \,\, {\cal L} \Big(\hspace*{-0.5mm}\mathbf{W}_k,P_j^*,t^*,\xi_k^*,\psi_j^*,\mathbf{Y}_k^*,\alpha_j^*,\theta_i^*\hspace*{-0.5mm}\Big).
\end{eqnarray}
Suppose $\mathbf{R}_k^*$ is not positive-definite. In this case, we can choose $\mathbf{W}_k=r\mathbf{w}_k\mathbf{w}_k^H$ as one of the optimal solution of (\ref{eqn:dual2}), where $r>0$ is a scaling parameter and $\mathbf{w}_k$ is the eigenvector corresponding to a non-positive eigenvalue $\rho_k \le 0$ of $\mathbf{R}_k^*$, i.e., $\mathbf{R}_k^*\mathbf{w}_k=\rho_k\mathbf{w}_k$.  We substitute $\mathbf{W}_k=r\mathbf{w}_k\mathbf{w}_k^H$ and $\mathbf{R}_k^*\mathbf{w}_k=\rho_k\mathbf{w}_k$ into (\ref{eqn:dual2}) which leads to
\begin{equation}\label{eqn:lagrangian_transformed}
\sum_{k=1}^K\Tr(\rho_k r\mathbf{w}_k\mathbf{w}_k^H)-r\sum_{k=1}^K\Tr\Big(\mathbf{w}_k\mathbf{w}_k^H
\big(\mathbf{Y}_k^*+\frac{\xi_k^*\mathbf{H}_k}{\Gamma^{\mathrm{DL}}_{\mathrm{req}_k}}\big)\Big)+\Delta.
\end{equation}
In (\ref{eqn:lagrangian_transformed}), the first term, $\sum_{k=1}^K\Tr(\rho_k r\mathbf{w}_k\mathbf{w}_k^H)$ is not positive. As for the second term, since the channel vectors of $\mathbf{g}_j$ and $\mathbf{h}_k$ are assumed to be statistically independent, and from (\ref{eqn:dual_variables}), we obtain $\sum_{k=1}^K\Tr\Big(\mathbf{w}_k\mathbf{w}_k^H
\big(\mathbf{Y}_k^*+\frac{\xi_k^*\mathbf{H}_k}{\Gamma^{\mathrm{DL}}_{\mathrm{req}_k}}\big)\Big) > 0$. By setting $r\rightarrow \infty$, the term $-r\sum_{k=1}^K\Tr\Big(\mathbf{w}_k\mathbf{w}_k^H \big(\mathbf{Y}_k^*+\frac{\xi_k^*\mathbf{H}_k}{\Gamma^{\mathrm{DL}}_{\mathrm{req}_k}}\big)\Big)\rightarrow -\infty$. In this case, the dual optimal value  becomes unbounded from below. However, the optimal value of the primal problem (\ref{eqn:sdp_relaxation}) is non-negative due to constraint $\widetilde{\text{C6}}$. Thus, strong duality cannot hold which leads to a contradiction. Therefore, $\mathbf{R}_k^*$ is a positive-definite matrix with probability one, i.e., $\Rank(\mathbf{R}_k^*)=N_{\mathrm{T}}$.
By applying (\ref{eqn:lagrangian_gradient}) and a basic inequality for the rank of matrices, we have
\begin{eqnarray}\label{eqn:Y greater than}
\hspace*{-3mm}&&\hspace*{-2mm}\Rank(\mathbf{Y}^*_k)+\Rank\big(\xi_k^*\frac{\mathbf{H}_k}{\Gamma^{\mathrm{DL}}_{\mathrm{req}_k}}\big)\\[-0cm]
\hspace*{-3mm} \notag&\ge &\hspace*{-2mm}\Rank\big(\mathbf{Y}^*_k+\xi_k^*\frac{\mathbf{H}_k}{\Gamma^{\mathrm{DL}}_{\mathrm{req}_k}}\big)=\Rank(\mathbf{R}_k^*)=N_\mathrm{T}\\[0.2cm]
\hspace*{-3mm}&\Rightarrow &\hspace*{-2mm} \Rank(\mathbf{Y}^*_k)\ge N_{\mathrm{T}}-1.\notag
\end{eqnarray}
Combining the inequalities in (\ref{eqn:rank less than}) and (\ref{eqn:Y greater than}), we obtain that $\Rank(\mathbf{Y}^*_k) = N_{\mathrm{T}}-1$. Thus, $\Rank(\mathbf{W}^*_k)=1$ holds with probability one.


\end{document}